%% file: OCTPaperBRJune29.tex
\documentclass[aps,pra,superscriptaddress,twocolumn]{revtex4-1}

\pdfoutput=1
\usepackage{graphicx}
\usepackage{dcolumn}
\usepackage{bm}
\usepackage{amssymb}
\usepackage{array}
\usepackage{amsmath}
\usepackage{natbib}
\usepackage[usenames, dvipsnames]{color}

\begin{document}

\title{High-Resolution Nanoscale Solid-State Nuclear Magnetic Resonance Spectroscopy}
\input{author_list.tex}

%\date{\today}

\begin{abstract}
We present a new method for high-resolution nanoscale magnetic resonance imaging (nano-MRI) that combines the high spin sensitivity of nanowire-based magnetic resonance detection with high spectral resolution nuclear magnetic resonance (NMR) spectroscopy. By applying NMR pulses designed using optimal control theory, we demonstrate a factor of 500 reduction of the proton spin resonance linewidth in a (50-nm)$^3$ volume of polystyrene, and image proton spins in one dimension with a spatial resolution below 2~nm.

%Magnetic resonance imaging (MRI) is a powerful non-invasive technique that has transformed our ability to study the structure and function of biological systems. Key to its success has been the unique ability to combine imaging with magnetic resonance spectroscopy. Although it remains a significant challenge, there is considerable interest in extending MRI spectroscopy to the nanometer scale because it would provide a fundamentally new route for determining the structure and function of complex biomolecules.
%We present data taken with a nanowire magnetic resonance force microscopy (MRFM) setup. We show how the capabilities of this very sensitive spin-detection system can be extended to include spectroscopy and nanometer-scale imaging by combining optimal control theory (OCT) techniques with magic echo sequences. We apply OCT-based dynamical-decoupling pulse sequences to nanoscale ensembles of proton spins in polystyrene, and demonstrate a 500-fold line-narrowing of the proton spin resonance, from $30 kHz$ to $60 Hz$. We further demonstrate 1-D imaging over a $35-nm$ region with an average voxel size of $2.2 nm$.

\end{abstract}

\pacs{}
\maketitle

\section{\label{sec:Intro}Introduction}

Magnetic resonance imaging (MRI) is a powerful non-invasive technique that has transformed our ability to study the structure and function of biological systems. Key to its success has been the unique ability to combine imaging with magnetic resonance spectroscopy. This capability has led to a host of different imaging modalities that rely on the ability to distinguish the interaction of spins with their local chemical environment. Although it remains a significant challenge, there is considerable interest in extending the capabilities of MRI spectroscopy to the nanometer scale (nano-MRI) because it would provide a fundamentally new route for determining the structure and function of complex biomolecules.

A key challenge for nano-MRI is having the ability to sense nanometer-scale ensembles of nuclear spins. In recent years, a number of different approaches to ultra-sensitive spin detection have been developed, including magnetic resonance force microscopy (MRFM) \cite{Rugar_2004,Degen_2009,Nichol_2013}, nitrogen-vacancy magnetometry \cite{Rugar_2014,Mamin_2013,Staudacher_2013}, and nanoscale superconducting quantum interference device (SQUID) magnetometry \cite{Vasyukov_2013}. These techniques have improved upon the spin detection sensitivity of conventional inductively-detected magnetic resonance by more than eight orders of magnitude, culminating in the detection of single electron \cite{Rugar_2004} and proton \cite{Sushkov_2014} spins, and MRI of tobacco mosaic virus particles with sub-10-nm spatial resolution \cite{Degen_2009}. As demonstrated by the virus MRI result, MRFM holds promise for studying biological samples in part due to its capability to image sufficiently long length scales, up to around 100 nm, with very fine resolution. To fully realize the potential of nano-MRI, however, we need a platform that combines ultra-sensitive spin detection with high-resolution magnetic resonance spectroscopy.

As the spin detection and imaging success of MRFM were being established, there were also efforts to demonstrate its utility for dipolar \cite{Degen_2005,Degen_2006}, quadrupolar \cite{Verhagen_2002}, and chemical-shift \cite{Eberhardt_2008} spectroscopy by employing magic echo pulse sequences and other techniques developed for conventional solid-state magnetic resonance imaging. These experiments were able to achieve spectral resolution under 1 kHz, showing the spectroscopy possibilities of MRFM, but were limited to relatively large spin ensembles with spatial resolution around 1 $\mu$m. As Eberhardt et al. write, the static gradient sources used in many MRFM experiments complicate spectroscopy attempts  \cite{Eberhardt_2008}. They solved this problem by moving the gradient source out of range of their sample during the encoding pulse sequence using a piezo.

In our MRFM experiments, we use a Current-Focusing Field Gradient Source (CFFGS) which generates large magnetic fields on nanometer length scales by focusing electric currents through a narrow ($\sim 100$ nm) metallic constriction. MRFM using a CFFGS has the potential to combine nanoscale imaging with magnetic resonance spectroscopy. The CFFGS can generate very large time-dependent magnetic fields ($\sim0.1$ T) and magnetic field gradients ($\sim 10^6$ T/m) over a wide frequency bandwidth (DC\,-\,GHz) \cite{Nichol_2012} in nanometer-scale volumes. As Nichol et al. demonstrated in a previous two-dimensional nano-MRI experiment, these characteristics provide the potential for very sensitive spin detection, high imaging resolution, and rapid spin manipulations on spin ensembles as large as (100 nm)$^3$ \cite{Nichol_2013}.

In the nano-MRI experiment by Nichol et al., the spatial resolution was determined by the strength of magnetic field gradients and the encoding time. The encoding time was limited by the $T_2^*$ spin dephasing time, dominated by homonuclear dipolar interactions between proton spins \cite{Nichol_2013}. To image with finer spatial resolution and to enable high-resolution spectroscopy, dipolar interactions between spins must be decoupled. The solid-state MRI community has developed a variety of techniques to acheive this. In particular, the magic echo pulse sequence has been used to decouple both homonuclear and heteronuclear interactions. In addition, it has long laboratory- and rotating-frame spin evolution periods, during which gradient pulses may be applied for spatial encoding. %\cite{Degen_2005,Eberhardt_2008}.

Recently, nanoscale solid-state chemical-shift spectroscopy has been performed using nitrogen-vacancy centers \cite{Aslam_2017}. In this work, the WAHUHA and MREV-8 pulse sequences were used to extend the spin coherence times. Such dynamical decoupling sequences typically require implementing the same unitary operation throughout the ensemble. While the CFFGS is capable of producing intense magnetic fields, the large $B_1$ variation over the sample volume makes the use of hard pulses unfeasible.
%One challenge to implementing these pulse sequences with a CFFGS is that our magnetic field source only generates fields with large gradients, making hard pulse rotations impractical.
Such experimental limitations can be efficiently dealt with using pulses designed with optimal control theory (OCT). Provided one has sufficient knowledge of the experimental conditions, such as resonance offsets, bandwidth limitations, and $B_1$ dispersion, OCT pulses can be designed to perform rapid, robust unitary transformations.

In this paper, we demonstrate rapid high-fidelity rotations in a nanoscale ensemble of proton spins by applying OCT-engineered pulses using the magnetic fields produced by the CFFGS. We present a new extension of optimal control theory (OCT) that incorporates Average Hamiltonian Theory (AHT) to allow for control of the average Hamiltonian over the duration of the pulse. We use this capability to perform dynamical decoupling during OCT pulses that have duration comparable to $T_2^*$. By incorporating OCT-AHT $\frac{\pi}{2}$ pulses into magic echo sequences, we demonstrate a factor of $500$ narrowing of the proton-resonance linewidth, from 29 kHz to 57 Hz, in a $(50$-$ \text{nm})^3$ volume of polystyrene. Taking advantage of this greatly extended coherence time, we perform one-dimensional Fourier imaging of proton spins in polystyrene with a spatial resolution between 1.8 nm\,-\,2.6 nm within a 35-nm region of the sample. This work serves as a demonstration of the power of OCT to achieve high-fidelity spin control for spectroscopy and imaging in CFFGS-based nano-MRI systems.

\section{\label{sec:App}Apparatus}
\begin{figure}
\includegraphics[scale=1]{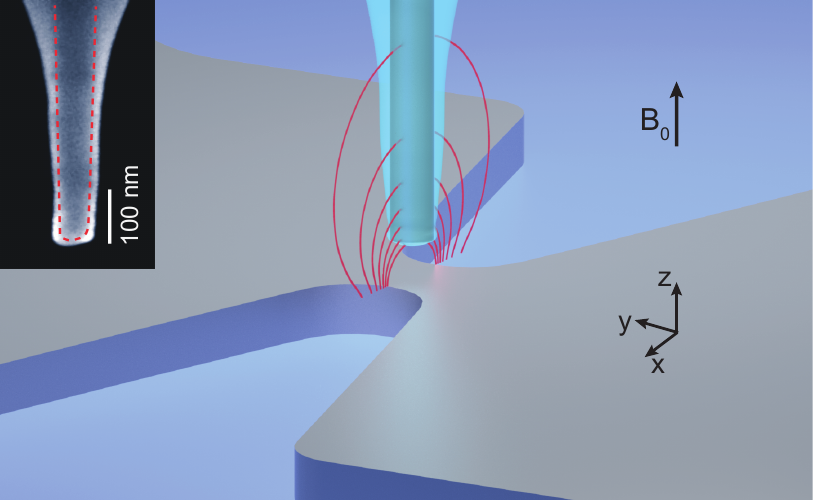}
\caption{\label{fig:App} Schematic rendering of the experimental apparatus, showing the polystyrene-coated nanowire, the Ag constriction device, and contour lines of constant Rabi frequency ($\omega_1$). The laser is focused to a roughly 3-$\mu$m diameter spot near the center of the SiNW. (Inset) Scanning electron micrograph of a representative SiNW tip coated with polystyrene. The outline of the nanowire is shown by the dashed line.}
\end{figure}

As shown in Fig. 1, we used an apparatus very similar to that described in our previous nano-MRI work \cite{Nichol_2013}. Force-detected magnetic resonance measurements were performed using silicon nanowire (SiNW) resonators, grown via the vapor-liquid-solid method. The SiNWs were modified using a procedure developed to split the frequencies of the two lowest-frequency flexural modes, while maintaining high quality factors. The details related to mode engineering of SiNW resonanators are discussed in the Supplemental Information (SI) section I. Finally, the SiNWs were coated at the tip with polystyrene dissolved in diethyl phthalate, using a micromanipulator. The modified SiNW resonator used for this experiment was approximately 20-$\mu$m long, with tip dimensions of 60 nm $\times$ 80 nm. At $4.2~\text{K}$, the two fundamental flexural modes had resonance frequencies $f_{1,2}=$ (315 kHz, 369 kHz), quality factors $Q_{1,2}=$ (8000, 8300), and spring constant $k\approx 1.0\times10^{-4}$ N/m.  We used a 2-$\mu$m wavelength laser interferometer polarized vertically along the length of the SiNW to measure its displacement \cite{Nichol_2008}. The lower-frequency mode was used because its oscillation direction was better aligned with the interferometer.

To generate time-varying magnetic fields and magnetic field gradients, we fabricated a CFFGS device by electron-beam lithography and liftoff of 10 nm/80 nm thick Ti/Ag films deposited on a MgO(100) substrate by electron-beam evaporation. The device contained a 150-nm wide and 100-nm long constriction, which served to focus electrical currents to produce the magnetic fields used for spin detection and control.

Spin detection was performed by flowing 70-mA peak current through the constriction near the SiNW resonance frequency, corresponding to a peak field gradient of $dB_z/dx=1.0\times 10^6$~ T/m near the tip of the SiNW. Additional information about the spin detection method is presented in SI section II. Magnetic resonance measurements were made at a proton resonance frequency of 48 MHz. The radio-frequency (RF) magnetic field used for spin control was produced by applying 50-mA peak current to the constriction, corresponding to a Rabi field $B_1=0.037$~T near the tip of the SiNW.

Fig. 2a shows a cross section of the proton Rabi frequency distribution at the center of the constriction. The magnetic fields produced by the constriction are calculated using the COMSOL Multiphysics finite-element simulation software. The simulations are based on the dimensions of the constriction measured using a scanning electron microscope.

The OCT pulses that were designed for these measurements occupy a wide bandwidth ($\sim10$ MHz). We found that waveform distortions caused by standing waves in the cables between the room temperature electronics and the CFFGS device can severly degrade the performance of the OCT pulses. We therefore developed a technique, based on force detection, that allowed us to characterize the complex transfer function of our electronics directly at the site of the spins. This capability was crucial for achieving good performance of the OCT pulses. The details of this technique will be provided in a separate publication. The spin-resonance frequency of 48 MHz was chosen to coincide with a relatively flat region of the transfer function.

The experiment took place in high vacuum at a base temperature of 4.2 K. We observed approximately 1-2 K heating of the nanowire due to the laser. Using Attocube nanopositioners and custom piezo walkers and scanners, the tip of the polystyrene-coated SiNW was positioned approximately 50-nm above the center of the constriction.

\begin{figure}
\includegraphics[scale=1]{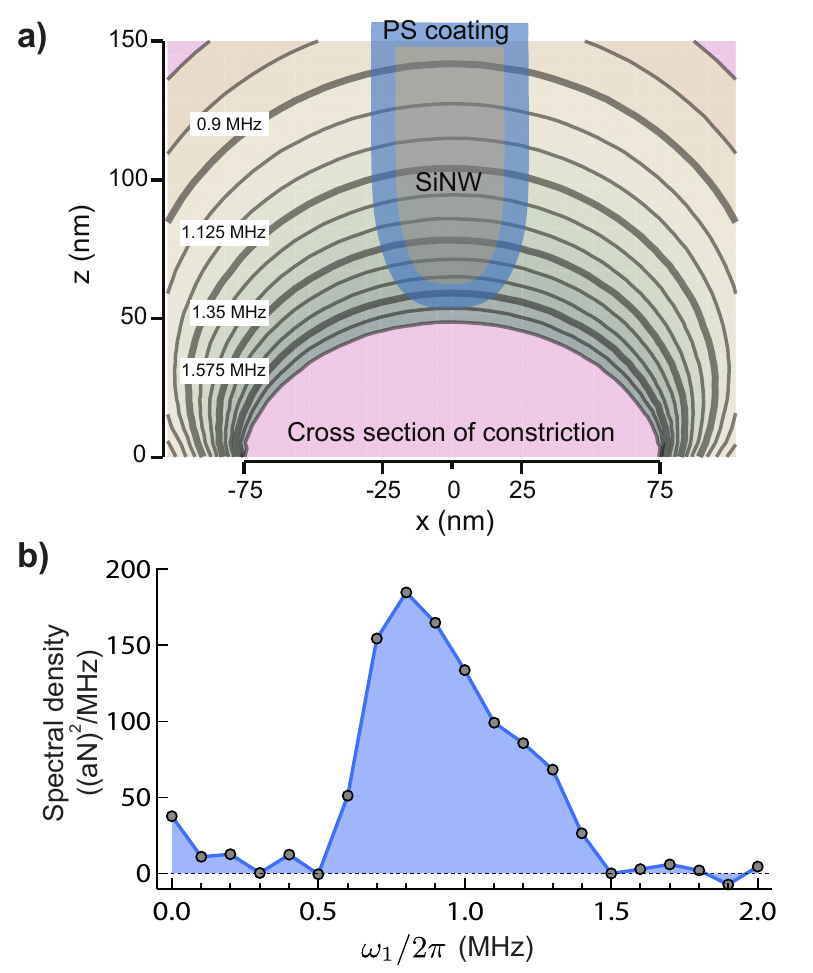}
\caption{\label{fig:RabiDisp} Proton Rabi frequency dispersion. (a) Schematic of polystyrene-coated nanowire and proton Rabi frequencies as a function of $x$ and $z$ positions relative to the center of the constriction. The Rabi frequencies are given by $\omega_1(x,0,z)/2\pi=\gamma B_x(x,0,z)/4\pi$ in the $xz$-plane centered on the constriction, where $\gamma$ is the proton gyromagnetic ratio (SI section III). Field values were calculated for 50-mA-pk RF current flowing through the constriction. (b) Spin signal spectral density as a function of proton Rabi frequencies $\omega_1/2\pi$. Spins near the tip of the SiNW, closest to the constriction, have the highest Rabi frequencies. The sharp falloff of the spectral density at lower frequencies is caused by the adiabatic passages used in the MAGGIC spin-detection protocol, which maintains adiabaticity for spins experiencing $\omega_1/2\pi \gtrsim 0.7$ MHz.}
\end{figure}

\section{\label{sec:OCT}Optimal Control Pulse Design} In order to implement most decoupling pulse sequences, one needs the ability to perform the same final spin rotation over the entire spin ensemble. Hard-pulse rotations are not achievable in our system due to the inhomogeneity of fields produced by the constriction. Certain state-to-state transfers such as adiabatic passages can be achieved easily even in the presence of an inhomogeneous $B_1$ field, but these are insufficient for decoupling sequences. OCT pulses have been used successfully in conventional NMR experiments to address various experimental constraints, including inhomogeneous $B_1$ fields.

The task of any OCT algorithm is to find a control sequence $a(t')$ over a period $0\le t' \le t$ that adheres to experimental constraints and generates the desired evolution for some ensemble of controlled systems. In the context of quantum control, $a(t')$ is generally an electromagnetic waveform, designed to generate a Hamiltonian evolution. Frequently, we are interested in a robust implementation of a particular unitary transformation $V$ for each member among the ensemble of interest; in this setting, we are dealing with an ensemble of nuclear spins. If $U_i \left[ a(t') \right]$ is the unitary operation generated on the $i^{th}$ spin at time $t$, we define a target function $\Phi \left(\lbrace U_i \rbrace\right)$, a functional of $a(t')$, that measures the total distance between $U_i \left[ a(t') \right]$ and $V$ for the entire ensemble. The central problem of OCT is to find an $a(t')$ that minimizes $\Phi \left[ a(t') \right]$. Many examples of numerical algorithms for solving this problem exist in the literature \cite{Khaneja_2005, Doria_2011}.

The benefit of OCT algorithms over attempting to find $a(t')$ analytically is at least twofold: analytical solutions (e.g. composite hard pulses) are usually restricted to a single system rather than an ensemble, and generally ignore deterministic experimental distortions to the waveform seen by the spins. Numerical pulse searches can easily accommodate both of these demands by incorporating a set of ensemble member specific transfer functions to the control search, which account for waveform distortions and differences in the Hamiltonian parameters. For OCT algorithms, the presence of such transfer functions merely adds an extra step into the computation and typically does not appreciably affect their performance. Robust control in the presence of $B_0$ and $B_1$ inhomogeneities \cite{Borneman_2010, Li_2011}, and compensation for waveform distortions \cite{Borneman_2012, Hinks_2015}, have been demonstrated for conventional NMR measurements.

In this work, we use the Gradient Ascent Pulse Engineering (GRAPE) algorithm \cite{Khaneja_2005}, which prescribes an efficient way to find the functional derivative $\delta\Phi[a(t')]/\delta a(t')$ of $\Phi$ with respect to $a(t')$ for piecewise constant control sequences $a(t')$. This converts the control search into a multivariate optimization problem. Given the functional derivative $\delta\Phi[a(t')]/\delta a(t')$ and a (usually) randomly generated initial guess $a_0(t)$, we use a simple multivariate gradient descent algorithm to reach the local minimum of $\Phi$ near $a_0(t)$. In practice, we generate a number of initial guesses until we find a waveform that yields a suitably low $\Phi$. The details of this procedure are further discussed in SI section IV. In general, the length of the control sequence will increase with the range of demands (e.g. range of targeted $B_0$ or $B_1$, bandwidth constraints for the waveform $a(t')$, etc.) required by the sequence. Additional demands typically require a larger number of initial guesses before a suitable sequence is found.

For this experiment, we searched for a pulse with a bandwidth of 10 MHz that would perform a $\frac{\pi}{2}$ rotation on all spins in a $B_1$ range, initially from 0.6-1.2 MHz, based on the experimental distribution of proton Rabi frequencies shown in Fig. 2b. Details of the Rabi frequency measurement are given in SI section III. Initially, we searched for a pulse that performed the desired unitary for this range of $B_1$, without explicitly averaging the dipolar Hamiltonian. The shortest pulse we found had a length of $t=7\ \mu$s, which is comparable to the measured $T_2^*=10.9\ \mu$s. To avoid dephasing over the duration of the pulse, we used a new method that incorporates Average Hamiltonian Theory (AHT) into the optimal control pulse design \cite{Haas_2017}. We added constraints into our pulse optimization to minimize the $0^{th}$-order Magnus terms of the homonuclear dipolar and chemical-shift Hamiltonians over the duration of the pulse. Because of these added constraints, the duration of the pulse increased from $7\,\mu s$ to $13\ \mu$s, however, the dephasing during the pulse was dramatically reduced. Details of the pulse optimization and calculated performance metrics are given in SI sections IV-VI.

\begin{figure}
\includegraphics[scale=1]{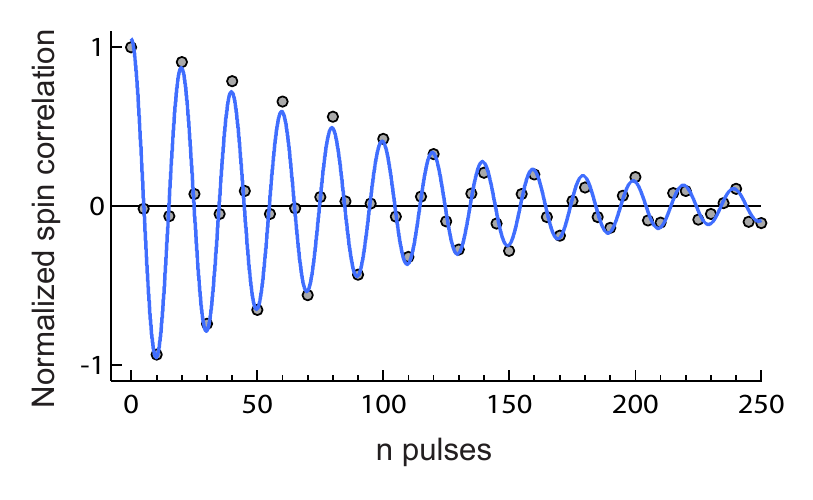}
\caption{\label{fig:RepPiHalf} Normalized spin correlation after repeated application of the $\Omega [\frac{\pi}{2}]_x$ pulse. The data is fit to the functional form $\cos(\pi n/10)\, e^{-cn}$. From the fit, we find $c=0.01$.}
\end{figure}

To experimentally test the performance of the 13-$\mu$s long OCT pulse, we applied repeated $\Omega [\frac{\pi}{2}]_x$ pulses to spins starting from the $z$-axis and measured the resulting $z$-axis magnetization. Here, $\Omega [\frac{\pi}{2}]_x$ refers to an OCT pulse that performs a $\frac{\pi}{2}]_x$ rotation. The data in Fig. 3 show the measured spin signal after sequential application of the $\Omega [\frac{\pi}{2}]_x$ pulse, in increments of 5 pulses. The signal decays to $e^{-1}$ after 100 pulses, indicating that the operation works with 99\% accuracy for spins starting from the $z$-axis.

This measurement serves as an initial demonstration of the power of OCT-AHT to produce high-fidelity unitary transformations tailored to our experimental conditions while averaging unwanted Hamiltonian terms to nearly zero. In this respect, these OCT pulses approximate delta-function hard pulses, making them perfectly suited to most NMR pulse sequences.

\section{\label{sec:SME}Magic Echoes}

\begin{figure*}
\includegraphics[scale=0.37]{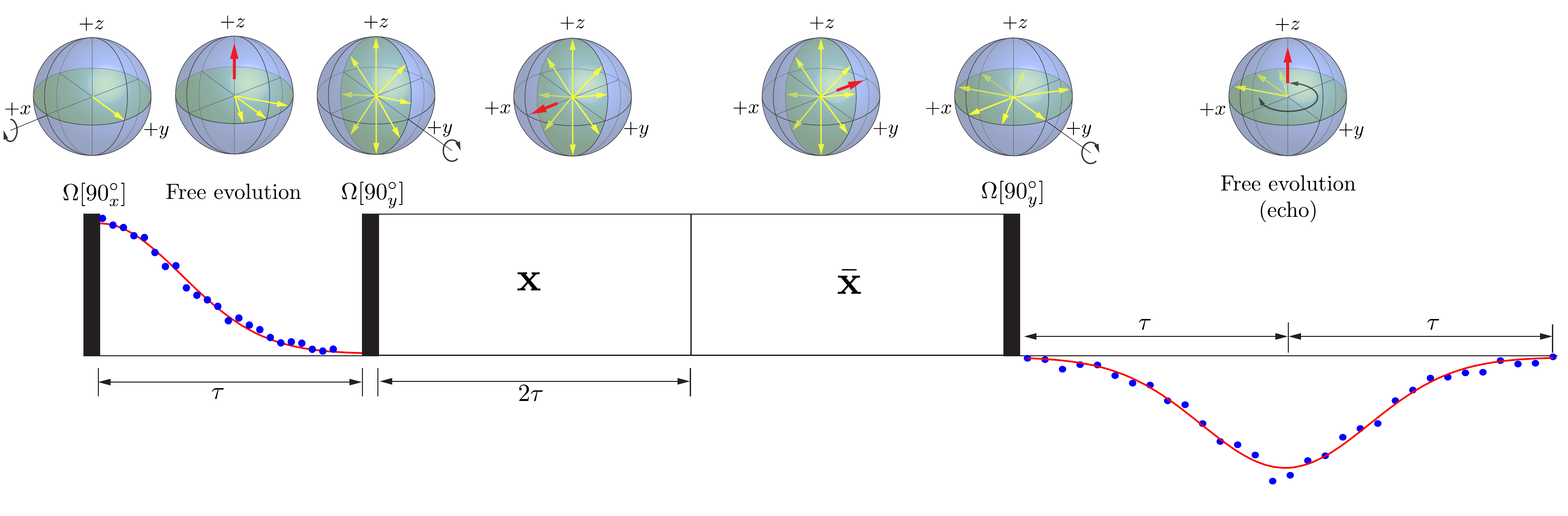}
\caption{\label{fig:ME} Pulse diagram and data (blue circles) for a magic echo pulse sequence performed with $\tau =25\ \mu$s, and with $13~\mu\text{s}$ OCT pulses in place of hard-pulse rotations. Bloch diagrams above show the spin rotations (yellow arrows) during the magic echo sequence. The red arrows point in the direction of the external magnetic field during free evolution periods, and in the direction of the effective field during the rotary echo.} %The echo amplitude is 87\% of the initial amplitude.
\end{figure*}

\begin{figure}
\includegraphics[scale=1]{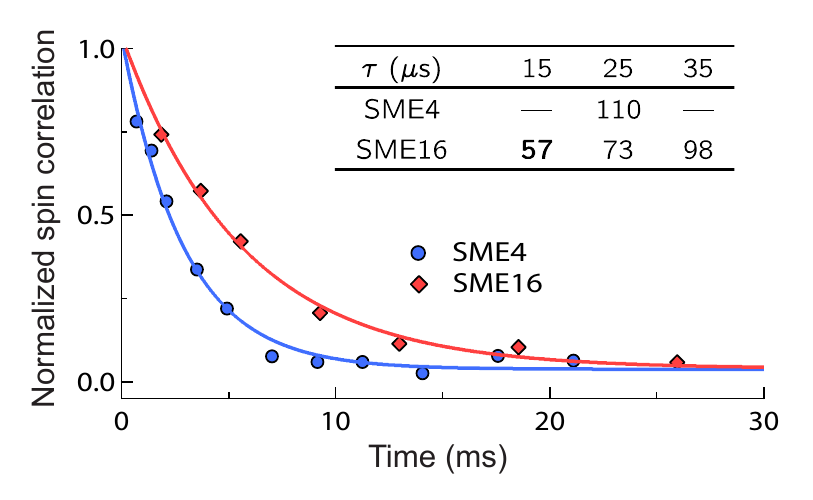}
\caption{\label{fig:SME} Normalized spin correlation vs. evolution time after sequential applications of an SME decoupling sequence. The evolution time includes the time during the OCT pulses. SME4 ($\tau = 25\ \mu$s) and SME16 ($\tau=15\ \mu$s) shown. (Inset) Table showing linewidths in Hz for sequential application of different SME sequences.}
\end{figure}

One of the most basic and important tests of the OCT pulses is to evaluate their performance as substitutes for hard-pulse rotations in dynamical decoupling pulse sequences. The magic echo is a basic decoupling pulse sequence frequently used in solid state NMR which refocuses the homonuclear dipolar Hamiltonian \cite{Hafner_1996}. In Fig. 4 we show data and the pulse diagram for a magic echo performed using $\Omega [\frac{\pi}{2}]_y$ rotations.

To perform the magic echo, we first created a coherence by applying an $\Omega [\frac{\pi}{2}]_x$ pulse. The spins then evolved under the static field $B_0$ for a time $\tau$. During this period, the spins dephased over a time $T_2^*$, primarily due to homonuclear dipolar coupling. After the first free evolution period $\tau$, an $\Omega [\frac{\pi}{2}]_y$ pulse was applied, followed by a rotary echo (RE) $(\mathbf{x}:\bar{\mathbf{x}})$ lasting a total time $4\tau$. The RE removes dephasing resulting from $B_1$ inhomogeneity. Following the RE, a second $\Omega [\frac{\pi}{2}]_y$ pulse was applied. After another free evolution period $\tau$, the echo formed as the total dipolar evolution was averaged to zero. In our implementation, we used two $\Omega [\frac{\pi}{2}]_y$ pulses, which also averaged the chemical-shift Hamiltonian to zero over the duration of the sequence, and caused an overall $\pi$ phase flip, resulting in the negative echo signal shown. %Magic echoes can also be applied that rescale the chemical shift to 0.3 in the delta-function limit.

Symmetric magic echo (SME) sequences are formed by combining several magic echoes around different axes \cite{Boutis_2003}. By doing so, the effect of certain pulse imperfections can be significantly reduced, resulting in longer coherence times than simply repeating a single magic echo. SME sequences have been demonstrated to have excellent line-narrowing capabilities, and can be used for chemical-shift spectroscopy and Fourier-transform imaging by encoding static ($B_0$) and radio-frequency ($B_1$) field strengths. We tested the SME4 and SME16 sequences, both of which remove chemical-shift effects, by applying the sequences repeatedly and observing the signal decay. Fig. 5 shows data sets obtained by repeated applications of the SME4 and SME16 pulse sequences. We observed that the SME16 sequence performed slightly better than the SME4, with the longest spin coherence time (5.6 ms) observed for SME16 ($\tau=15\,\mu$s). For all measurements, the value of $\tau$ was chosen to be longer than $T_2^*$ to avoid spin-lock effects \cite{Matsui1992,Matsui1995}.

\section{\label{sec:Imaging}Imaging}
\begin{figure*}
\includegraphics[scale=1]{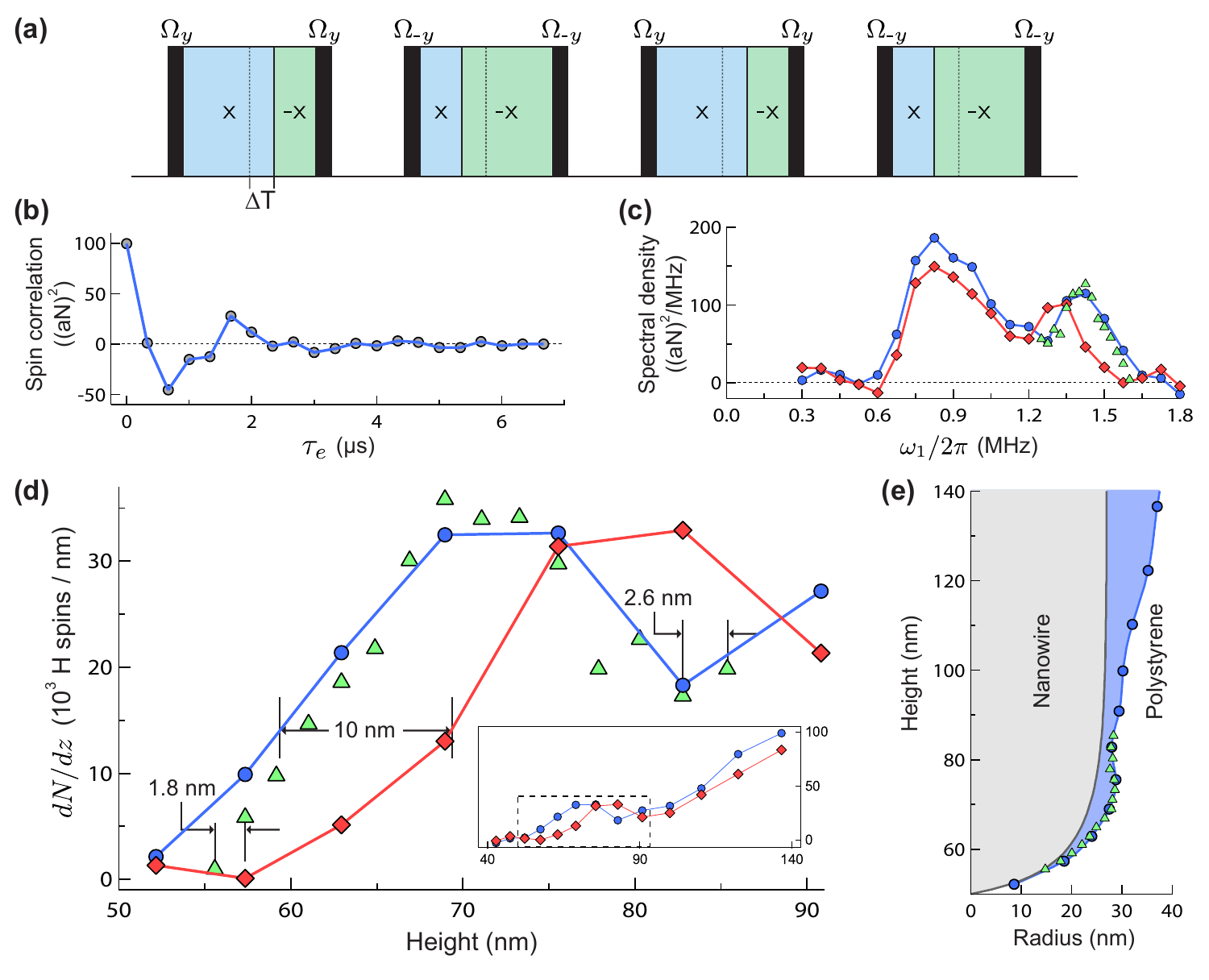}
\caption{\label{fig:Imaging} One-dimensional imaging.  (a) Modified SME4 spin sequence used for one-dimensional imaging includes time offsets ($\Delta T$) during rotary evolution periods to encode Rabi frequency information.  (b) Spin signal after a single modified SME4 as a function of total rotary evolution time $\tau_e$. Data shown were taken with 0.33-$\mu$s steps and with a tip-surface separation of 55 nm.  (c) Spectral density of spin signal as a function of $\omega_1/2\pi$, obtained by performing discrete cosine transforms on time-domain data. Green data taken with 55-nm tip-surface separation and 25-kHz frequency steps; blue data taken with 55-nm tip-surface separation and 75-kHz frequency steps; red data taken with 65-nm tip-surface separation and 75-kHz frequency steps. (d) One-dimensional proton density as a function of distance from the constriction. Tip-surface separation and frequency step size are the same for each data set as those in part (c). (e) Model of the SiNW tip with polystyrene coating constructed using the $^1$H-spin density for polystyrene $4.9\times 10^{28}$~m$^{-3}$, and assuming a cylindrically symmetric nanowire and polystyrene coating.}
\end{figure*}

The long spin evolution periods within the magic echo sequence are ideal for imaging. Spatial information can be encoded by applying $B_0$ gradient pulses during the two laboratory-frame evolution periods lasting $2\tau$, and $B_1$ gradients during the rotating-frame evolution lasting $4\tau$. In this experiment, we offset the rotary evolution times, while fixing the total rotary evolution period to be $4\tau$, to encode spatial information using the large $B_1$ inhomogeneity produced by the CFFGS.

To test the high-resolution imaging capabilities of our system, we applied an SME4 sequence with the RE portion modified for $B_1$ encoding (Fig. 6a). With the modified RE, the spins evolve in the $B_1$ gradient for a time $\tau_e=8 \Delta T$ within a single SME4 sequence, where $\Delta T$ is the offset introduced to each RE. To verify that the introduction of an offset into the RE did not degrade the refocusing properties of the SME4, we applied two back-to-back modified SME4 sequences, changing the sign of $\Delta T$ between the SME4 sequences to undo the evolution in the $B_1$ gradient. We compared the spin signal after applying the two modified SME4 sequences to the spin signal after applying two SME4 sequences that were not modified. The magnitude of the signal was the same for the two cases, confirming that the modified RE maintained the refocusing properties of the original SME4.

For the imaging measurements, a $7.5$-$\mu$s long OCT pulse was designed to target spins experiencing maximum Rabi frequencies from 0.9-1.75 MHz, which allowed us to accurately measure the spins closest to the constriction. These new pulses were used in the SME4 sequence to collect the imaging data shown in Fig. 6. To image the one-dimensional spin distribution as a function of Rabi frequency, we incremented $\Delta T$ and measured the corresponding spin correlation. Fig. 6b shows the correlation as a function of $\tau_e$. To find the distribution of spins as a function of $\omega_1$, we performed a cosine transform on the time-domain data (Fig. 6c). Note that the frequency resolution of the green data (25 kHz) is much finer than that seen in Fig. 2b (100 kHz), which was taken without the benefit of the SME4 sequence.

To convert the frequency-domain data to a function of position, we used the simulated magnetic field data shown in Fig. 2a. We took the $B_1$ field at each vertical position to be the average over a $50\times 50$ (nm)$^2$ region in the $x$ and $y$ dimensions. Since the spin sensitivity of our measurement depended on the gradient of $B_1$, we also used this magnetic field simulation to convert the measured spin correlation into a number of detected spins, giving the spin density plot shown in Fig. 6d. To validate the image reconstruction, we physically retracted the tip by 10 nm, from 55-nm to 65-nm above the surface of the constriction. After doing so, the calculated spin density reconstructed 10-nm further above the surface (Fig. 6d), confirming our model of the magnetic fields produced by the constriction.

Finally, by assuming an average SiNW radius of 27~nm, and a rotationally symmetric distribution of polystyrene on the circumference of the SiNW, we reconstructed a one-dimensional image of the polystyrene coating on the tip of the SiNW, shown in Fig. 6e.

\begin{table}
\caption{\label{tab:EncodingRes} Spatial encoding using a single SME16 with $\tau=15 \ \mu$s.}
\begin{ruledtabular}
\begin{tabular}{c c c c}
&Encoding&Encoding&Encoding\\[-1pt]
&gradient ($\partial B$)\footnotemark[1]&time ($\tau_e$)\footnotemark[2]&limit ($\lambda$)\footnotemark[3]\\
\hline\noalign{\smallskip}
$B_1$ encoding&3 G/nm&960 $\mu$s&0.4 \AA\\
$B_0$ encoding&9 G/nm&480 $\mu$s&0.3 \AA\\
\end{tabular}
\end{ruledtabular}
\begin{flushleft}
\footnotemark[1]{Average gradients in a region 50\,-\,80 nm above the constriction produced by applying $50$-$\text{mA-pk}$ ($B_1$-encoding) and $70$-$\text{mA-pk}$ ($B_0$-encoding) current through the CFFGS.} \\
\footnotemark[2]{Maximum encoding times for an SME16 are 64$\tau$ ($B_1$-encoding) and 32$\tau$ ($B_0$-encoding).}\\
\footnotemark[3]{Spatial encoding limits are calculated as $\lambda=\pi / \gamma\, \partial B\, \tau_e$ where $\gamma$ is the proton gyromagnetic ratio, $\partial B$ is the average encoding gradient, and $\tau_e$ is the maximum encoding time.}
\end{flushleft}
\end{table}

\section{Conclusion}

We have demonstrated the ability to encode the one-dimensional spin distribution with an average resolution of 2 nm over a 30-nm vertical region. To achieve this resolution, we used only a small fraction of the total encoding time available to us. Table I shows the available encoding times during a single SME16 ($\tau =15\ \mu$s). We see that with the gradients and encoding times available to us, it is possible to encode spins with sub-Angstrom resolution.

This capability opens up the possibility for NMR ``diffraction'' measurements, first proposed by Mansfield and Grannell in 1973 \cite{Mansfield_1973}. In one dimension, NMR diffraction works by matching the encoding wavenumber $k=\gamma G \tau_e$ to an integer multiple of the reciprocal lattice vector $2\pi/a$, where $a$ is the lattice constant, $\gamma$ is the spin gyromagnetic ratio, $G$ is the magnetic field gradient, and $\tau_e$ is the encoding time. The extension to three dimensions is straightforward. By analogy to Bragg scattering, in NMR diffraction a coherent signal is produced from a macroscopic ensemble of spins only if the Bragg condition is satisfied.

Atomic scale NMR diffraction has not been demonstrated, because until now the combination of long coherence times and the large magnetic field gradients necessary to encode spins at the Angstrom scale was not available. NMR diffraction would provide a fundamentally new way to study the atomic structure of crystalline solids by providing chemically-specific structure factor information. In particular, NMR diffraction could be very useful for the study of biological systems that possess periodic structures.

In this work, we have used the intense magnetic fields produced by a CFFGS in combination with OCT pulses that incorporate AHT to realize high-fidelity spin control in nanometer-scale ensembles of nuclear spins. By combining this capability with high-sensitivity spin detection, we have demonstrated high-resolution solid-state spectroscopy and imaging in a system of proton spins with strong spin-spin interactions. We believe that the combination of these capabilities is very powerful and will enable new approaches in nano-MRI.

\section{Acknowledgements}
This work was supported by the US Army Research Office through Grant No. W911NF-16-1-0199 and by the Department of Physics at the University of Illinois, and was carried out in part in the Frederick Seitz Materials Research Laboratory Central Research Facilities at the University of Illinois. D.G.C. would like to acknowledge Canadian Excellence Research Chairs (CERC). R.B. and D.G.C. would like to acknowledge the support of the Canada First Research Excellence Fund (CFREF), the Natural Sciences and Engineering Research Council of Canada (NSERC), the Canadian Institute for Advanced Research, the Province of Ontario and Industry Canada. R.B. would like to thank Ben Yager for  performing ESR measurements of paramagnetic defects in silicon nanowires.

\bibliography{bib}{}

\end{document}

\title{Supplementary Information for High-Resolution Nanoscale Solid-State Nuclear Magnetic Resonance Spectroscopy}
\maketitle
\input{measurementContent}

\input{pulseFindingContent}
\pagebreak
\bibliography{bib}{}

%% file: author_list.tex
\author{William Rose}
\altaffiliation{W. Rose and H. Haas contributed equally to this work.}
\affiliation{Department of Physics, University of Illinois at Urbana-Champaign, 1110 W. Green St., Urbana, IL 61801-3080, USA}
\author{Holger Haas}
\altaffiliation{W. Rose and H. Haas contributed equally to this work.}
\affiliation{Department of Physics, University of Waterloo, Waterloo, ON, Canada, N2L3G1}
\affiliation{Institute for Quantum Computing, University of Waterloo, Waterloo, ON, Canada, N2L3G1}
\author{Angela Q. Chen}
\affiliation{Department of Physics, University of Illinois at Urbana-Champaign, 1110 W. Green St., Urbana, IL 61801-3080, USA}
\author{Nari Jeon}
\affiliation{Department of Materials Science and Engineering, Northwestern University, Evanston, Illinois 60208, USA}
\author{Lincoln J. Lauhon}
\affiliation{Department of Materials Science and Engineering, Northwestern University, Evanston, Illinois 60208, USA}
\author{David G. Cory}
\affiliation{Institute for Quantum Computing, University of Waterloo, Waterloo, ON, Canada, N2L3G1}
\affiliation{Department of Chemistry, University of Waterloo, Waterloo, ON, Canada, N2L3G1}
\affiliation{Perimeter Institute for Theoretical Physics, Waterloo, ON, Canada, N2L2Y5}
\affiliation{Canadian Institute for Advanced Research, Toronto, ON, Canada, M5G1Z8}
\author{Raffi Budakian}
\email{rbudakian@uwaterloo.ca}
\affiliation{Department of Physics, University of Illinois at Urbana-Champaign, 1110 W. Green St., Urbana, IL 61801-3080, USA}
\affiliation{Department of Physics, University of Waterloo, Waterloo, ON, Canada, N2L3G1}
\affiliation{Institute for Quantum Computing, University of Waterloo, Waterloo, ON, Canada, N2L3G1}
\affiliation{Perimeter Institute for Theoretical Physics, Waterloo, ON, Canada, N2L2Y5}
\affiliation{Canadian Institute for Advanced Research, Toronto, ON, Canada, M5G1Z8}

%% file: measurementContent.tex
\section{\label{sec:NWSplitting} Nanowire Mode splitting, passivation, and sample attachment}
We used silicon nanowires (SiNWs) grown by the vapor-liquid-solid method with drop-cast gold nanoparticle catalysts. The nanowires were grown to be about 20-$\mu$m long, with a slight taper at the base, and a tip diameter of roughly 120 nm. The wires' radial symmetry makes the two lowest-frequency flexural modes nearly degenerate. The SiNW used for the experiment initially had fundamental modes $f_{1,2}=$ 617.7 kHz, 618.6 kHz with $Q=23,000$ for both modes at room temperature. For the spin measurements, we needed to couple to only one of the flexural modes. Therefore, we developed a procedure to separate the frequencies of the fundamental modes by removing radial symmetry.

\begin{figure}
\includegraphics[scale=1]{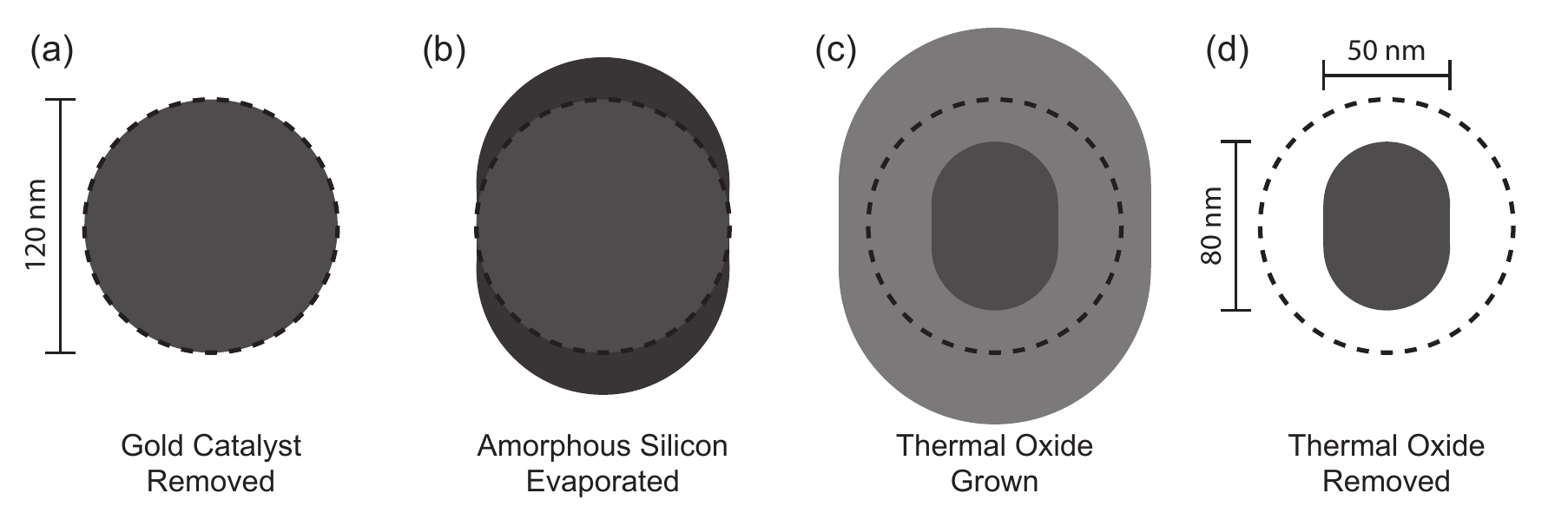}
\caption{\label{fig:NWPrep} Nanowire-shaping diagram showing the approximate shape of the nanowire cross section at four stages of fabrication. The dashed line indicates the initial nanowire shape.}
\end{figure}

Some of the processes required the SiNW to be exposed to high temperature. To prevent diffusion of the Au catalyst, we first used an ion mill to remove the Au catalyst by orienting the nanowire tips towards the ion source. We used a bias voltage of 100 V because the low voltage increases the milling selectivity of Au over Si to about 10:1. We had previously calibrated the milling rate, and confirmed with a scanning electron microscope that the gold particles were fully removed. Fig. \ref{fig:NWPrep}a shows a diagram of the cross section of the nanowire after the Au catalyst removal.

Next, we used a electron-beam evaporator to directionally deposit high-purity silicon on the top and bottom of the nanowire (Fig. \ref{fig:NWPrep}b). We deposited approximately a 20-nm thick Si layer on both the top and bottom of the SiNW at a rate of 0.2 \AA/s. By depositing equal amounts of Si on both sides, we prevented the nanowire from warping due to differential stress produced by the amorphous Si layer. This deposition step removed the nanowire’s radial symmetry, separating the frequencies of the two lowest-frequency flexural modes. However, the amorphous Si layer dramatically lowers the quality factor to $Q\approx 1,000$.

To improve the quality factor, we next grew oxide to consume the amorphous evaporated silicon. We used a tube furnace with 1 atm pure oxygen at 900 $^\circ$C to consume approximately 60 nm of Si from the diameter of the SiNW (Fig. \ref{fig:NWPrep}c). Finally, the thermal oxide was removed using vapor hydrofluoric acid, which very selectively etches SiO$_2$ but not Si, leaving behind a smaller, crystalline, assymetric nanowire (Fig. \ref{fig:NWPrep}d). The resultant nanowire had tip dimensions of about 50$\times$80 nm, $f_{1,2}=$ 362 kHz, 423 kHz with $Q\approx 7,500$ for both modes at room temperature. 

Next, we grew and passivated a thin thermal oxide layer on the SiNW to try to reduce the concentration of dangling bonds and defects on the surface. Oxide quality is improved at higher growth temperatures. To grow a thin layer of oxide at high temperature, we preheated a furnace tube with 1 atm O$_2$ to 1000 $^\circ$C, then gradually inserted the silicon nanowire chip over 30 s, left it in the furnace for 20 s, and gradually removed it over another 30 s. Based on calibration measurements made on Si wafers, this procedure should have grown about 3-4 nm of silicon oxide. Immediately after the oxide growth, we moved the nanowire chip to another tube furnace and annealed it for 2 hours at 300 $^\circ$C in 1 atm of 5\% H$_2$/Ar forming gas.

To attach the polystyrene sample, we dissolved 5200 MW polystyrene in diethyl phthalate, which has a very low vapor pressure (0.002 mmHg at room temperature). We coated the tip of a micropipette with a small drop of the solution, and used high precision manual translation stages under an optical microscope with a long working distance to carefully insert the tip of the target nanowire approximately 1-2 $\mu$m into the polystyrene solution. Using a scanning electron microscope would degrade the polystyrene, therefore, initial confirmation of polystyrene attachment was performed by measuring the nanowire resonance frequencies. The polystyrene mass-loads the nanowire, lowering the resonance frequencies to $f_{1,2}=$ 313 kHz, 367 kHz at room temperature. Final confirmation that there is polystyrene on the last 100 nm of the nanowire, where we can measure it, comes from the magnetic resonance images of the spin distribution.

\section{\label{sec:ModMAGGIC}Modified MAGGIC Protocol}

\begin{figure}
\includegraphics[scale=1]{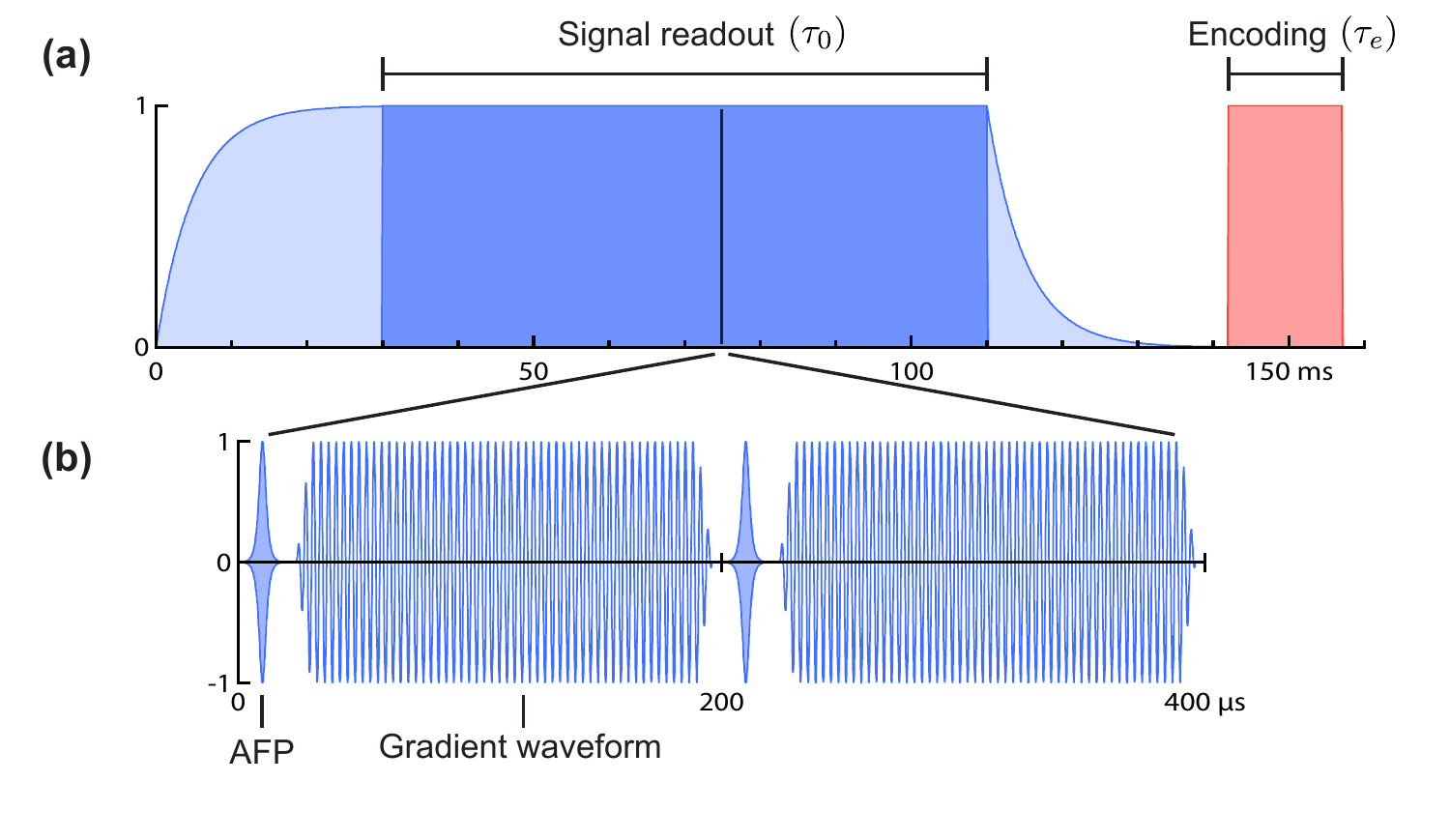}
\caption{\label{fig:ModMag} Modified MAGGIC protocol diagram. (a) Approximate timing diagram for single modified MAGGIC block. Light blue regions indicate the gradient modulation ramps. Dark blue areas indicate the readout period during which the adiabatic inversions and gradient modulation are applied. (b) Detailed timing diagram showing the adiabatic inversions and gradient waveform applied during the signal readout period.}
\end{figure}

\begin{figure}
\includegraphics[scale=.7]{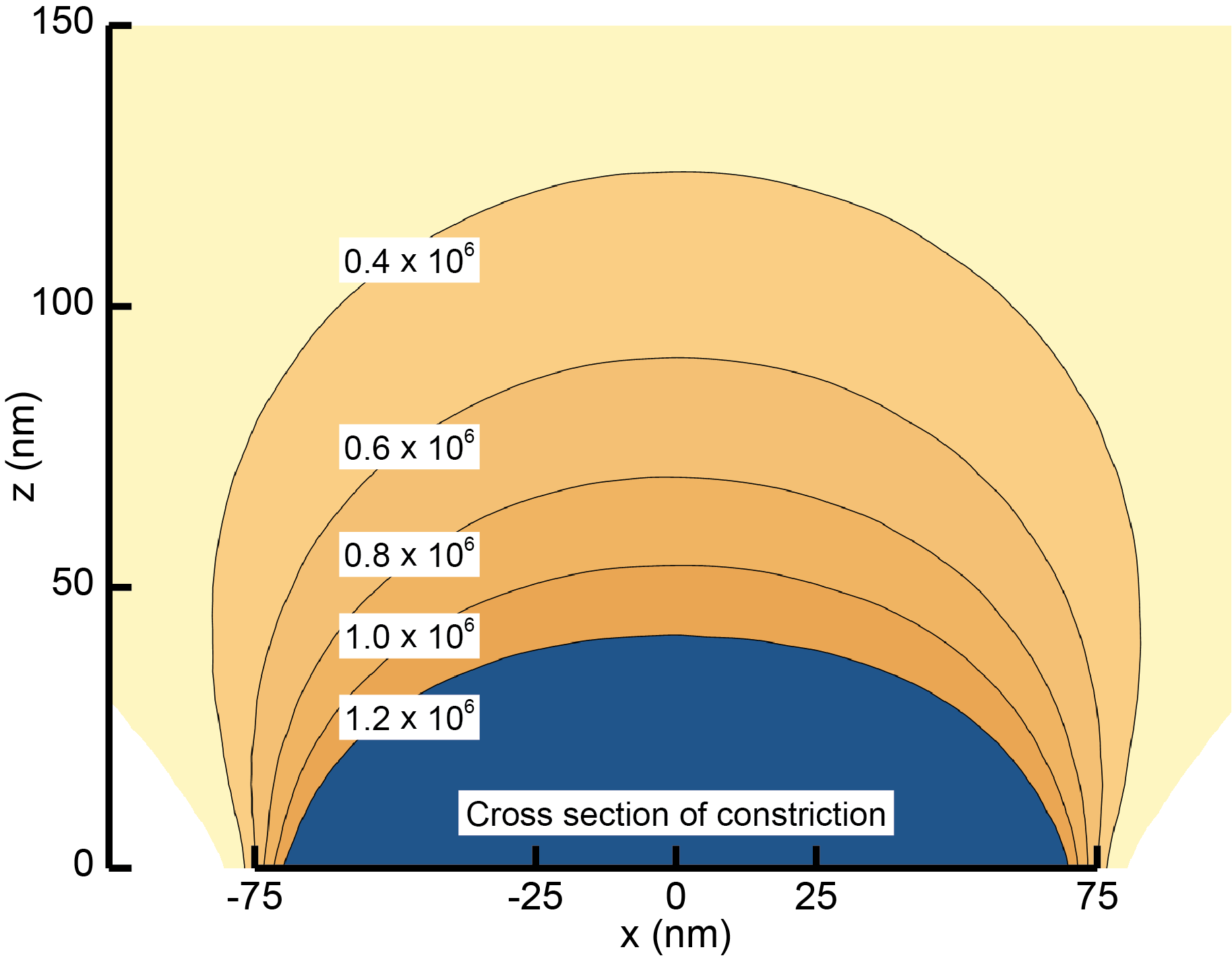}
\caption{\label{fig:Gradient} Contour map of $dBz/dx$ gradients (T/m) calculated for 70-mA-pk current passing through the constriction. These gradients are generated at the nanowire resonance frequency during the readout periods of the MAGGIC protocol, and are those referred to as the gradient modulation and denoted by $G(\vec{r})$ in the equations below. The center of the top surface of the constriction is at the position (0,0).}
\end{figure}

We measure the $\sqrt{N}$ fluctuations in an ensemble of $\sim10^6$ proton spins. Because the mean spin signal is zero, we measure the average correlation between two consecutive measurements, where the duration of each measurement is chosen to be much shorter than the correlation time of the statistical fluctuations. The magnetic resonance pulse sequence is applied between the two spin-measurement blocks.

For all the measurements described in this paper, we measured the spin signal using a modified version of the MAGGIC protocol described by Nichol et al. in 2012 \cite{Nichol_2012}. The overall pulse scheme, shown in Fig. \ref{fig:ModMag}a, consisted of a readout period, during which the MAGGIC protocol was applied at maximum amplitude, and an encoding period during which the MAGGIC protocol was off. The encoding sequence consisted of an SME sequence or other spin manipulations neccessary for a particular measurement. The MAGGIC protocol was ramped on and off exponentially with time constant $\tau=\frac{2Q}{\omega_0}= 8.1$ ms, where $\omega_0=2\pi \times 315$ kHz is the angular frequency of the fundamental resonance mode of the nanowire, and $Q=8000$ is the quality factor of the fundamental mode. There is electrostatic coupling between the SiNW and the electric fields that drive the gradient modulation currents through the CFFGS. The exponential ramp allows the nanowire to ring up and down gradually when these fields are turned on and off, minimizing transients. Adiabatic spin inversions were not applied during the ramp times to minimize the decay of the statistical spin correlations.

To minimize the electrostatic coupling between the SiNW and the gradient modulation electric fields, we applied a DC bias voltage to the CFFGS to null the static charge on the tip of SiNW resulting from a potential difference between the SiNW and the CFFGS. We found that the electrostatic coupling was minimized by applying a potential difference of $\sim 0.6$ V between the CFFGS and the SiNW. In addition, feed-forward cancellation was used to minimize the sidebands caused by the gradient modulation \cite{Nichol_2012}. We experimented with the exact timings for this sequence, and settled on a ramp time of $\tau_{ramp}=30$ ms, and a readout time of $\tau_0=80$ ms. The encoding time $\tau_e$ was varied between tens of microseconds and several milliseconds depending on the desired encoding sequence.

Fig. \ref{fig:ModMag}b shows the MAGGIC protocol block, consisting of two short hyperbolic secant adiabatic full passages (AFPs), and two periods of sinusoidal gradient modulation at the nanowire resonance frequency $f_0$. The second gradient-modulation block turns on with the same phase as the first at time $t=(n+\frac{1}{2})t_0$, where n is a positive integer (we used $n=63$), and $t_0$ is the period of the nanowire resonance. This ensures that there is no Fourier component of the gradient modulation at the nanowire resonance frequency.

Following the MAGGIC protocol theory developed by Nichol et al. \cite{Nichol_2012}, during the readout periods, a single spin produces a force at the nanowire resonance frequency $f_0$ with average amplitude $F_0 = \mu D G$. $\mu$ is the magnetic moment of the spin, $D$ is the duty cycle of the $f_0$ gradient waveform, and $G=\frac{dB_z}{dx}$ is the peak magnetic field gradient at the location of the spin (Fig. \ref{fig:Gradient}). The longitudinal correlation of each spin is Markovian, described by the random telegraph function $h(t)\in \{\pm 1\}$, with the correlations $\int_0^\infty h(t')h(t'+\tau)dt'=e^{-\tau/\tau_m}$. In our measurements we determined $\tau_m=0.6$ s.

Now let us consider the correlation function for a single spin at position $\vec{r}$. The signal obtained from such a spin will be the average correlation of the $\hat{z}$ spin measurement made before and after the encoding period:

\begin{equation}
C(\vec{r},U)=\Theta(\vec{r},U)\frac{1}{2}\frac{\mu^2 G^2\!(\vec{r})D^2}{\tau_0^2}\left\langle \int_{0}^{\tau_0} h(\vec{r},t')dt' \int_{\tau_0}^{2\tau_0} h(\vec{r},t)dt \right\rangle,
\end{equation}

where  $\left\langle...\right\rangle$ refers to the average of an ensemble of measurements, and

\begin{equation}
\Theta(\vec{r},U)=\left\langle \uparrow \right| U^\dagger\!(\vec{r})\, \sigma_z\, U (\vec{r}) \left| \uparrow \right\rangle,
\end{equation}

where $U(\vec{r})$ is the unitary transformation applied to a spin at position $\vec{r}$ during the encoding time $\tau_e$. In writing SI, we have assumed that the encoding time is much shorter than the longitudinal relaxation time $(\tau_e\ll T_1)$. We will consider how the random telegraph noise for a single spin affects the correlation measurement.

\begin{equation}
\left\langle \int_{0}^{\tau_0} h(\vec{r},t')dt' \int_{\tau_0}^{2\tau_0} h(\vec{r},t)dt \right\rangle=\int_{0}^{\tau_0} dt' \int_{\tau_0}^{2\tau_0} dt \left\langle h(\vec{r},t') \, h(\vec{r},t) \right\rangle,
\end{equation}

with

\begin{equation}
\left\langle h(\vec{r},t') \, h(\vec{r},t) \right\rangle=e^{-|t-t'|/ \tau_m}.
\end{equation}

Now we can perform the integration:

\begin{equation}
\int_{0}^{\tau_0} dt' \int_{\tau_0}^{2\tau_0} dt \  e^{-|t-t'|/ \tau_m}=\tau_m^2 \left(1-e^{-\tau_0 / \tau_m}\right)^2.
\end{equation}

Returning to (S1), the average correlation for a single spin is given by:

\begin{equation}
C(\vec{r},U)=\Theta(\vec{r},U)\frac{1}{2}\frac{G^2(\vec{r})D^2}{\tau_0^2}\, \tau_m^2 \left(1-e^{-\tau_0 / \tau_m}\right)^2.
\end{equation}

We treat each spin in the measurement ensemble as statistically independent. This allows us to express the total correlation as:
\begin{equation}
C(U)=\frac{1}{2}D^2 \left( \frac{\tau_m}{\tau_0} \right)^{\! 2} \left(1-e^{-\tau_0 / \tau_m}\right)^{\! 2} \! \int \rho(\vec{r}) G^2(\vec{r}) \Theta (\vec{r}, U) d\vec{r},
\end{equation}

where $\rho (\vec{r})$ is the spin density and the integration is done over the measured volume.

\section{\label{sec:RabiMeas}Rabi measurement}

\begin{figure}
\includegraphics[scale=0.5]{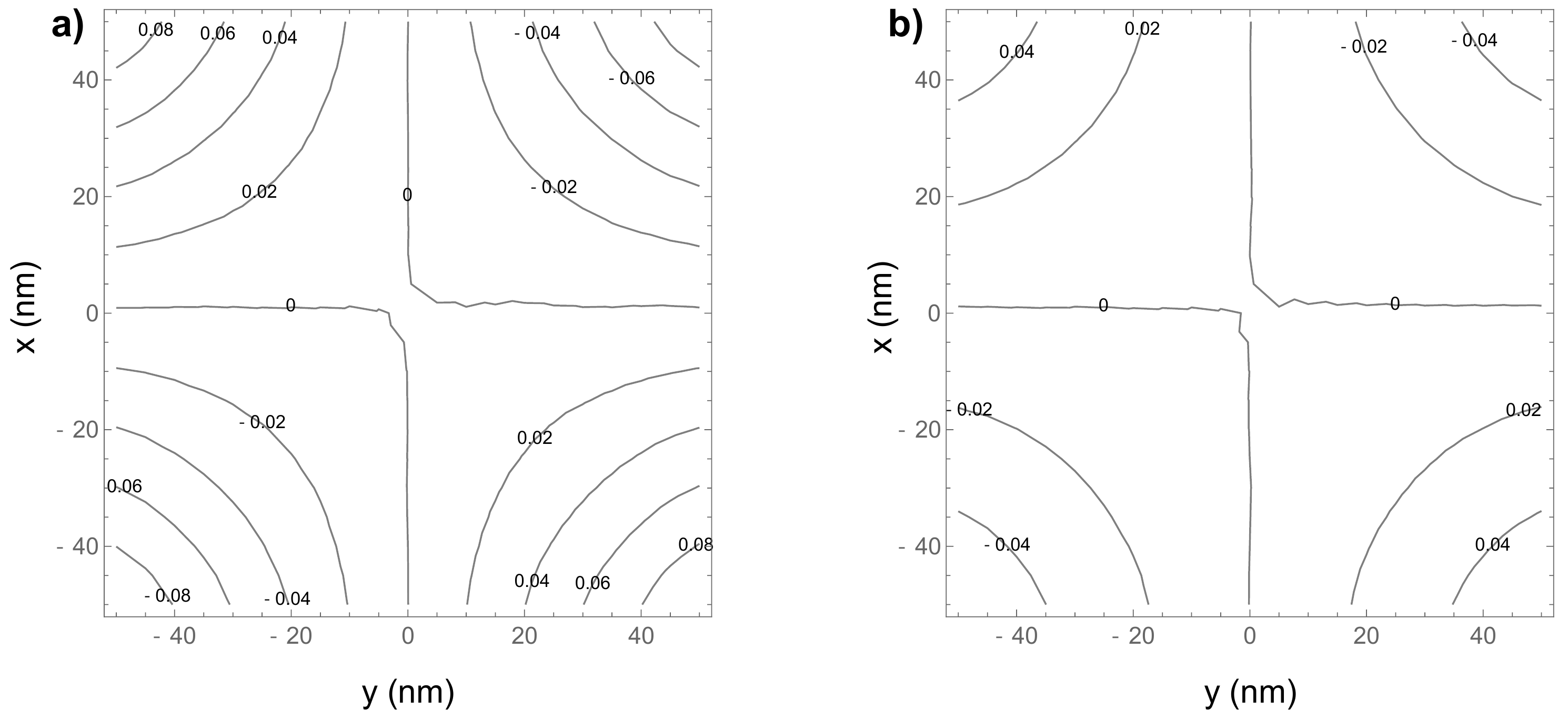}
\caption{\label{fig:By_Bx_ratio} The ratio of $B_y$ and $B_x$ generated by the constriction in a plane parallel to the constriction surface. a) $B_y \left( x, y, z = 50~\text{nm} \right) / B_x \left(x, y, z = 50~\text{nm} \right)$ b) $B_y \left( x, y, z = 100~\text{nm} \right) / B_x \left(x, y, z = 100~\text{nm} \right)$, where $z$ denotes the height measured from the constriction and the origin corresponds to the center of the constriction.}
\end{figure}

To measure the distribution of Rabi frequencies over our spin sample, we applied hard pulses around $\hat{x}$ to the spins. This produced a field-dependent rotation, giving the unitary transformation:
\begin{equation}
U(\vec{r})=e^{-i\sigma_x t_p\omega_1\!(\vec{r})/2},
\end{equation}
where $t_p$ is the duration of the pulse, and $\omega_1(\vec{r})=\gamma B_x(\vec{r}) / 2$ is the position-dependent Rabi frequency defined in the main text. It may be seen in Fig. \ref{fig:By_Bx_ratio} that within our sample volume $B_y(\vec{r})$ is at least an order of magnitude smaller than $B_x(\vec{r})$. Since $B_1(\vec{r})=\left(B_x^2(\vec{r})+B_y^2(\vec{r})\right)^{1/2}/2$ and $B_y^2(\vec{r})\ll B_x^2(\vec{r})$ we used the simplifying approximation that $B_1(\vec{r})=B_x(\vec{r})/2$.

We define:
\begin{equation}
\theta(\vec{r})\equiv t_p \omega_1\!(\vec{r}).
\end{equation}

To avoid dephasing errors, we kept $t_p$ constant while varying the amplitude of $\omega_1(\vec{r})$. However, we were interested in the distribution of Rabi frequencies for the maximum applied current through the constriction (50 mA). Therefore we define new variables, the position-dependent Rabi frequency at maximum current: $\omega_{1max}(\vec{r})$, and an effective time $t_e$ such that:

\begin{equation}
\theta(\vec{r})=t_e~ \omega_{1max}(\vec{r}).
\end{equation}

For the initial Rabi distribution measurement shown in Fig. 2b, we took steps corresponding to a $t_e$ increment of 0.25 $\mu$s from $t_e=0\ \mu$s to $t_e=5\ \mu$s, and measured the spin correlation at each step. We performed a discrete cosine transform (DCT-I) on these data to produce the spin correlation as a function of $\omega_{1max}(\vec{r})$. The transformed data correspond to 0.1 MHz steps of $\omega_{1max}/2\pi$ from 0\,-\,2.0 MHz.

%% file: pulseFindingContent.tex
\section{\label{sec:PulseSearch}Pulse Search}
Pulse search techniques described in \cite{Haas_2017} provided us with a capability to simultaneously engineer arbitrary unitary operations and selectively suppress evolution under unwanted system Hamiltonian terms by averaging their respective Average Hamiltonian Theory (AHT) \cite{Waugh_1968} terms to zero. Crucially, we could do this over a range of control parameters (e.g. Rabi frequencies and resonance offsets) and in the presence of known deterministic distortions to the control sequence. For this work we engineered control sequences that implemented $\pi/2$ spin rotations and suspended the evolution under homonuclear dipolar interactions ($D=\mathbb{\sigma}.\mathbb{\sigma}-3~\sigma_z \otimes \sigma_z$) and $\sigma_z$ Hamiltonians. Accordingly, our pulse optimization target function combined three terms: fidelity to the target unitary, $V = \exp \left(-i \frac{\pi}{2} \frac{\sigma_x}{2} \right)$,  and the operator norms of zeroth order AHT terms for $\sigma_z$ and dipolar Hamiltonians.

We parametrized our control Hamiltonian as
\begin{align}
   \label{eq:controlHamiltonian}
   H_\textit{ctrl} (t, \gamma) = \gamma~ \Omega(t) \left( \cos[\phi(t)] \frac{\sigma_x}{2} + \sin[\phi(t)] \frac{\sigma_y}{2} \right) ,
\end{align}
where we took the amplitude modulation function $\Omega(t)$ to be a positive unitless function $0 \le \Omega(t) \le 1$, and the phase modulation function $\phi(t)$ to satisfy $-\pi \le \phi(t) \le \pi$. The Rabi strength parameter $\gamma$ took values between $2 \pi \nu_\text{min}$ and $2 \pi \nu_\text{max}$, with $\nu_\text{min}$ and $\nu_\text{max}$ being respectively the minimum and maximum targeted Rabi strengths. Over a period $0 \le t_1 \le t$ the control Hamiltonian $H_\textit{ctrl} (t, \gamma)$ generates a unitary
\begin{align}
  \label{eq:fidelity}
  U(t,\gamma) = \mathbb{T} \exp \left[-i \int_0^t dt_1~ H_\textit{ctrl} (t_1, \gamma) \right],
\end{align}
where $\mathbb{T}$ denotes the time ordered exponential. Zeroth-order AHT prescribes setting the following integrals to zero: $I_\textit{D}(\gamma) = \int_0^t dt_1 \left[ U^\dagger(t_1,\gamma) \otimes U^\dagger(t_1,\gamma) \right]. D. \left[ U(t_1,\gamma) \otimes U(t_1,\gamma) \right]$ to suspend the dipolar interaction and $I_{\sigma_z}(\gamma) = \int_0^t dt_1~ U^\dagger(t_1,\gamma) . \sigma_z. U(t_1,\gamma)$ to eliminate chemical shift and resonance offset effects for a particular $\gamma$ value.

The three quantities minimized during the pulse engineering are the unitary metric $F_U(\gamma)$, dipolar metric $F_D(\gamma)$, and $\sigma_z$ metric  $F_{\sigma_z}(\gamma)$, defined as
\begin{align}
   \label{eq:unitaryMetric}
   F_U(\gamma) &= \frac{1}{2} \sqrt{ \left| \text{Tr} \left[ e^{i \frac{3 \pi}{2} \frac{\sigma_x}{2}}. U(t,\gamma) \right] \right|^2 + \left| \text{Tr} \left[ \sigma_y^\dagger . U(t,\gamma) \right] \right|^2 + \left| \text{Tr} \left[ \sigma_z^\dagger . U(t,\gamma) \right] \right|^2 } , \\
  \label{eq:dipolarMetric}
    F_D(\gamma) &= \sqrt{\frac{\text{Tr} \left[ I_\textit{D}^\dagger(\gamma).I_\textit{D}(\gamma) \right]}{\text{Tr} \left[ I_\textit{D}^\dagger(0).I_\textit{D}(0) \right]}}, \\
    \label{eq:chemicalMetric}
    F_{\sigma_z}(\gamma) &= \sqrt{\frac{\text{Tr} \left[ I_{\sigma_z}^\dagger(\gamma).I_{\sigma_z}(\gamma) \right]}{\text{Tr} \left[ I_{\sigma_z}^\dagger(0).I_{\sigma_z}(0) \right]}} .
\end{align}
The quantities (\ref{eq:unitaryMetric}), (\ref{eq:dipolarMetric}) and  (\ref{eq:chemicalMetric}) only take values between 0 and 1 and are identically zero for a control sequence $\lbrace \Omega_x(t), \Omega_y(t) \rbrace = \lbrace \Omega(t) \cos \left[ \phi(t) \right], \Omega(t) \sin \left[ \phi(t) \right] \rbrace$ which for a particular $\gamma$ yields $U(t,\gamma) = V$ while setting $I_D(\gamma)$ and $I_{\sigma_z}(\gamma)$ to zero. Following the method in \cite{Haas_2017, Khaneja_2005} we divided the pulse length $t$ into intervals of equal length $\Delta t = \frac{t}{N}$, $N$ being the number of steps, and work with piecewise constant controls $\lbrace \Omega_x \left( t \right), \Omega_y \left( t \right) \rbrace = \lbrace \Omega_x^{(i)}, \Omega_y^{(i)} \rbrace$ for $i \Delta t \le t < (i+1) \Delta t$ and $i \in \left\lbrace 0,1, ..., N-1 \right\rbrace$. Consequently, the pulse metrics $F_U(\gamma)$, $F_D(\gamma)$ and $F_{\sigma_z}(\gamma)$ became functions of $\lbrace \Omega_x^{(i)}, \Omega_y^{(i)} \rbrace$ and $\gamma$, with their partial derivatives, e.g. $\left\lbrace \frac{\partial}{\partial \Omega_x^{(j)}} F_U \left(\lbrace \Omega_x^{(i)}, \Omega_y^{(i)} \rbrace, \gamma \right) , \frac{\partial}{\partial \Omega_y^{(j)}} F_U \left(\lbrace \Omega_x^{(i)}, \Omega_y^{(i)} \rbrace, \gamma \right) \right\rbrace$, easily evaluated by matrix exponential methods introduced in \cite{Haas_2017}. We minimized the cost function
\begin{align}
   \label{eq:costFunction}
   \Phi \left(\lbrace \Omega_x^{(i)}, \Omega_y^{(i)} \rbrace \right) = \frac{1}{14} \sum_{\gamma_i \in \Gamma} \left[ \frac{5}{9} F_U \left(\lbrace \Omega_x^{(i)}, \Omega_y^{(i)} \rbrace, \gamma_i \right) + \frac{3}{9} F_D \left(\lbrace \Omega_x^{(i)}, \Omega_y^{(i)} \rbrace, \gamma_i \right) + \frac{1}{9} F_\sigma \left(\lbrace \Omega_x^{(i)}, \Omega_y^{(i)} \rbrace, \gamma_i \right) \right]
\end{align}
using a standard gradient descent algorithm, $\Gamma$ being a set of fourteen $\gamma$ values distributed roughly uniformly over the Rabi range $2\pi \nu_\text{min}$ and $2\pi \nu_\text{max}$. The relative weights for the individual pulse metrics in (\ref{eq:costFunction}) were picked so to give approximately equal minimization rates for $F_U$, $F_D$ and $F_{\sigma_z}$. All searches were started with $\left\lbrace \Omega_x^{(i)} \right\rbrace$ and $\left\lbrace \Omega_y^{(i)} \right\rbrace$ drawn from independent uniform pseudorandom distributions over a range $\left[ -\frac{1}{\sqrt{2}}, \frac{1}{\sqrt{2}} \right]$. We kept increasing the pulse length until we observed sufficiently rapid convergence speeds. The final pulse for $\nu_\text{min} = 0.6~\text{MHz}$, $\nu_\text{max} = 1.2~\text{MHz}$ turned out to be $13$-$\mu\text{s}$ long with $N=522$, and for $\nu_\text{min} = 0.9~\text{MHz}$, $\nu_\text{max} = 1.75~\text{MHz}$ we found a $7.5$-$\mu\text{s}$ long pulse with $N=360$.

\begin{figure}
\includegraphics[scale=0.5]{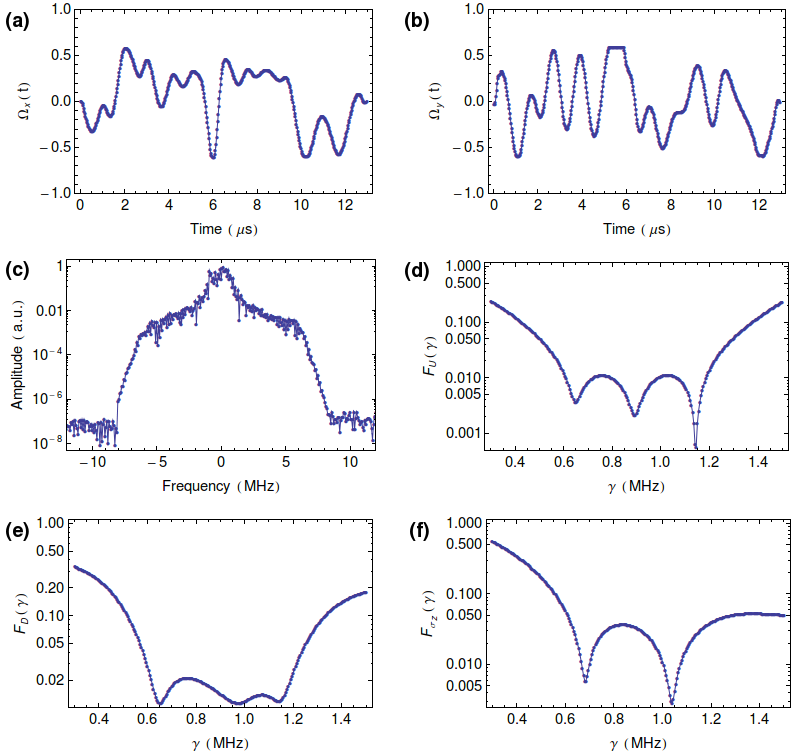}
\caption{\label{fig:13usPulse} (a) In phase amplitude $\Omega_x(t)$ of the 13-$\mu\text{s}$ pulse. (b) Quadrature amplitude $\Omega_y(t)$ of the 13-$\mu\text{s}$ pulse. (c) Absolute value of the Fourier transformed pulse centred at the carrier frequency showing the limited spectral range of the pulse. (d) Unitary metric $F_U(\gamma)$ defined in (\ref{eq:unitaryMetric}) as a function of Rabi strength parameter $\gamma$, it can be seen that the pulse targets the range $2\pi \cdot 0.6~ \text{MHz}$ to $2\pi \cdot 1.2~ \text{MHz}$. (e) Dipolar metric $F_D(\gamma)$ defined in (\ref{eq:dipolarMetric}) as a function of Rabi strength parameter $\gamma$. (f) $\sigma_z$ metric $F_{\sigma_z}(\gamma)$ defined in (\ref{eq:chemicalMetric}) as a function of Rabi strength parameter $\gamma$.}
\end{figure}

Furthermore, to minimize transients, we fixed the first and last five steps of the pulse to have zero amplitude. We also incorporated a combined transfer function of a numerical bandpass filter and experimentally determined amplitude and phase transfer functions to account for pulse distortions due to our radio frequency (RF) electronics. This procedure is described in the next section. The resulting waveform and plots of the individual metrics for the $13$-$\mu\text{s}$ long pulse are plotted in Fig. \ref{fig:13usPulse}.

\section{\label{sec:Transfer}Transfer Function}
\begin{figure}
  \includegraphics[scale=0.5]{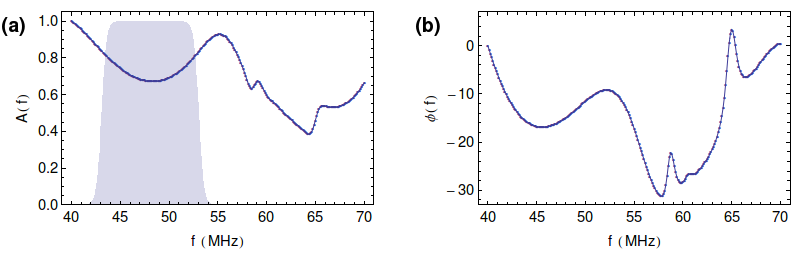}
 \caption{\label{fig:TransferFn} (a) Experimentally determined amplitude transfer function $A(f)$ as a function of carrier frequency $f$, the shaded area depicts the numerical bandpass filter used $A_\text{filter} \left(f,48~\text{MHz},10~\text{MHz} \right)$ defined in Equation (\ref{eq:bandPass}). (b) Experimentally determined phase transfer function $\phi(f)$.}
\end{figure}

We verified that our RF electronics exhibited a linear response over the range of output power used in our measurements. Therefore all distortions to the pulse due to the electronics could be characterized by the amplitude and phase transfer functions of the system, $A(f)$ and $\phi(f)$, respectively. We measured $A(f)$ and $\phi(f)$ of our electronics over a frequency range of $40~\text{MHz} \le f \le 70~\text{MHz}$ which are presented in Fig. \ref{fig:TransferFn}. We used this transfer function in conjunction with a numerical bandpass filter to limit the frequency range of the waveform by multiplying $A(f)$ with a numerical bandpass filter $A_\text{filter}(f)$ given by the function
\begin{equation}
  \label{eq:bandPass}
  A_\text{filter}(f,f_0,\Delta f) = \frac{1}{4} \left( 1 + \tanh \left[ \frac{20}{\Delta f} \left(f - f_0 + \frac{\Delta f}{2} \right) \right] \right) \left( 1 - \tanh \left[ \frac{20}{\Delta f} \left(f - f_0 - \frac{\Delta f}{2} \right) \right] \right) ,
\end{equation}
where $f_0$ denotes the carrier frequency and $\Delta f$ the bandwidth of the filter. This ensured smooth cut-off of frequencies outside of $\left[f_0 - \frac{\Delta f}{2}, f_0 + \frac{\Delta f}{2} \right]$. The numerical filtering was done to ensure that the spectral range of the pulse remained in the region where the amplitude transfer function was relatively flat and the phase transfer function relatively linear, thereby reducing distortions to the waveform caused by errors in determining the transfer function. We used $f_0=48~\text{MHz}$ and $\Delta f=10~\text{MHz}$ yielding a numerical amplitude transfer function shown in Fig. \ref{fig:TransferFn}, while the limited spectral range of the 13-$\mu\text{s}$ pulse can be seen in Fig. \ref{fig:13usPulse}c.

\section{\label{sec:Sims}Simulations}
To demonstrate the necessity for including AHT into our optimization we have compared the performance of the 13-$\mu\text{s}$ pulse presented in Fig. \ref{fig:13usPulse} with a pulse optimized without the averaging targets. Despite the latter being roughly half the length of the averaging pulse it performed substantially worse in numerical simulations of an eight spin network. For the simulations, we evaluated the Hamiltonian evolution of eight spins under dipolar couplings and various Rabi strengths and resonance offsets; these simulations provided a good way to assess the performance of pulses, since the effects of experimental distortions and transients was totally removed, and the ability to change the strength of various Hamiltonian terms enabled us to easily discern between different contributors (e.g. Rabi dispersion, resonance offesets/chemical shifts, dipolar evolution) to pulse performance. This benefit did come with a severe limitation to the system size, hence we could only expect the coherence times found in eight spin simulations to put an upper bound to the experimental results.

\begin{itemize}
\item {\bf Single Spin Simulations:} We performed single spin and eight spin simulations when analysing our pulses, which were represented as two-dimensional arrays $\lbrace \Omega_x^{(i)}, \Omega_y^{(i)} \rbrace$, $i \in \lbrace 1,...,N \rbrace$. Single spin simulations for a pulse with $N$ steps of length $\Delta t$ were carried out by evaluating the unitary ($U_\textit{I} (\gamma)$) at the end of the pulse for a set of $\gamma$ values sampling the Rabi distribution. The unitary is given by
\begin{align}
  \label{eq:1spinUnitary} 
  U_\textit{I} (\gamma) = e^{-i \gamma \left( \Omega_x^{(N)} \frac{\sigma_x}{2} + \Omega_y^{(N)} \frac{\sigma_y}{2} \right) \Delta t} ...~ e^{-i \gamma \left( \Omega_x^{(2)} \frac{\sigma_x}{2} + \Omega_y^{(2)} \frac{\sigma_y}{2} \right) \Delta t} e^{-i \gamma \left( \Omega_x^{(1)} \frac{\sigma_x}{2} + \Omega_y^{(1)} \frac{\sigma_y}{2} \right) \Delta t} .
\end{align}
The simulated signal $\left\langle \sigma_z \right\rangle_\textit{I}$ after $n$ applications of the pulse was calculated as a weighted average of $\sigma_z$ expectation values over an experimentally determined probability distribution $\text{prob}(\gamma)$ given in Fig. \ref{fig:Distributions},
\begin{align}
  \label{eq:1spinSignal} 
  \left\langle \sigma_z \right\rangle_\textit{I}(n) = \sum_{\gamma_i \in \Gamma} \text{prob}(\gamma_i) \left\langle \uparrow \right| \left[ U_\textit{I}^\dagger(\gamma_i) \right]^n \sigma_z \left[ U_\textit{I} (\gamma_i) \right]^n \left| \uparrow \right\rangle .
\end{align}

\begin{figure}
\includegraphics[scale=0.5]{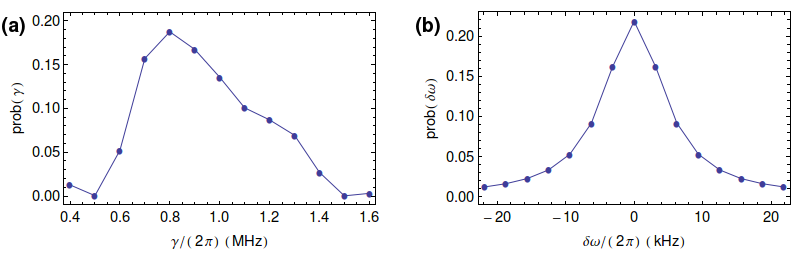}
\caption{\label{fig:Distributions} (a) Experimentally determined distribution of Rabi strengths. (b) $10.6~\text{kHz}$ wide Lorentzian distribution corresponding to the measured field inhomogeneity sampled at 15 points. We used the points and their respective probabilities in our spin signal simulations, with the probability distributions normalized such that $\sum_{\gamma_i \in \Gamma} \text{prob}(\gamma_i) = 1$ and $\sum_{\delta \omega_j \in \Delta} \text{prob}(\delta \omega_j) = 1$.}
\end{figure}

\item {\bf Eight Spin Simulations:} The eight spin simulations were carried out using a nuclear spin dipolar network representing the atomic structure of styrene molecule (polystyrene monomer). The atomic positions $\left\lbrace \vec{r_{i}} \right\rbrace$ for the eight protons in a single molecule were retrieved from Wolfram Mathematica Chemical Data database \cite{Mathematica}, ignoring the $^{13}$C nuclei. The dipolar Hamiltonian for the spin network in a particular field orientation $\hat{\zeta}$ is given by
\begin{align}
  \label{eq:dipolarHamiltonian}
   H_\textit{dipolar}(\hat{\zeta}) = \sum_{i=1}^{7} \sum_{j=i+1}^{8} \frac{\mu_0 \gamma^2 \hbar^2}{4 \pi \left| \vec{r_{ij}} \right|^3} \frac{1}{2} \left( 1 - 3 \left[ \frac{\vec{r_{ij}}.\hat{\zeta}}{\left| \vec{r_{ij}} \right|} \right]^2 \right) \frac{1}{4} \left( \mathbb{\sigma}^{(i)}.\mathbb{\sigma}^{(j)} - 3~\sigma_z^{(i)} \otimes \sigma_z^{(j)} \right) ,
\end{align}
where $\hat{\zeta}$ is a unit vector pointing along the direction of external magnetic field $B_0$, and $\vec{r_{ij}} = \vec{r_{i}} - \vec{r_{j}}$ is a vector connecting a pair of nuclei. Since our experimental results indicated a broad distribution of resonance offsets, we also included a corresponding term  $\delta \omega$ into our simulations. The eight spin unitaries for a pulse $\lbrace \Omega_x^{(i)}, \Omega_y^{(i)} \rbrace$  were functions of $\gamma$, $\hat{\zeta}$ and $\delta \omega$, and were evaluated as
\begin{align}
  \label{eq:8spinUnitary}
  U_\textit{VIII} (\gamma,\delta \omega,\hat{\zeta}) = e^{-i \gamma \left[ \Omega_x^{(N)} J_x + \Omega_y^{(N)} J_y + \delta \omega J_z + H_\textit{dipolar}(\hat{\zeta}) \right] \Delta t} ...~ e^{-i \gamma \left[ \Omega_x^{(1)} J_x + \Omega_y^{(1)} J_y + \delta \omega J_z + H_\textit{dipolar}(\hat{\zeta}) \right] \Delta t}
\end{align}
with $J_x = \sum_{i=1}^8 \frac{\sigma_x^{(i)}}{2}$, $J_y = \sum_{i=1}^8 \frac{\sigma_y^{(i)}}{2}$ and $J_z = \sum_{i=1}^8 \frac{\sigma_z^{(i)}}{2}$.

We calculated the spin signal $\left\langle 2 J_z \right\rangle_\textit{VIII}$ after $n$ applications of the pulse as a weighted average over experimentally determined distributions of resonance offsets $(\delta \omega)$ and Rabi strengths $(\gamma)$ given in Fig. \ref{fig:Distributions}. Furthermore, because our polystyrene sample was not a single crystal we included an equiprobable average of 50 random orientations $\lbrace \hat{\zeta_k} \rbrace$ sampling the surface of a unit sphere into the definition of
\begin{align}
  \label{eq:8spinSignal}
  \left\langle 2 J_z \right\rangle_\textit{VIII}(n) = \sum_{\gamma_i \in \Gamma} \sum_{\delta \omega_j \in \Delta} \sum_{\hat{\zeta_k} \in \Xi} \frac{\text{prob}(\gamma_i)~\text{prob}(\delta \omega_j)}{50} \left\langle \uparrow \right|^{\otimes 8} \left[ U_\textit{VIII}^\dagger(\gamma_i, \delta \omega_j, \hat{\zeta_k}) \right]^n 2J_z \left[ U_\textit{VIII}(\gamma_i, \delta \omega_j, \hat{\zeta_k}) \right]^n \left| \uparrow \right\rangle^{\otimes 8}.
\end{align}

\begin{figure}
\includegraphics[scale=0.5]{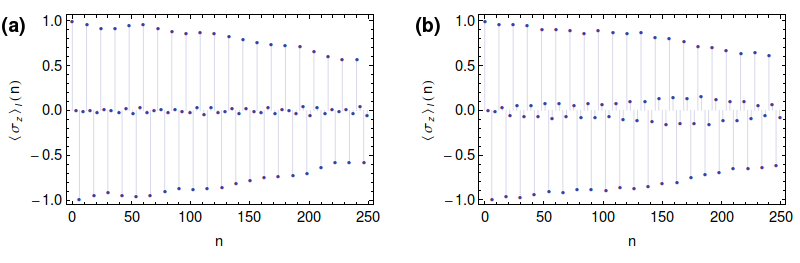}
\caption{\label{fig:1SpinSim} (a) Single spin simulation results of $\left\langle \sigma_z \right\rangle_\textit{I}(n)$ defined in (\ref{eq:1spinSignal}) for the 13-$\mu\text{s}$ pulse optimized with AHT terms and (b) simulations of $\left\langle \sigma_z \right\rangle_\textit{I}(n)$ for the 7-$\mu\text{s}$ pulse optimized without AHT terms as a function of $n$, the number of pulses applied.}
\end{figure}

\begin{figure}
\includegraphics[scale=0.5]{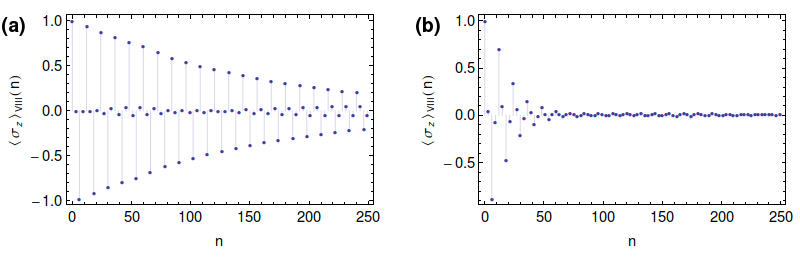}
\caption{\label{fig:8SpinSim} (a) Eight spin simulation results of $\left\langle 2 J_z \right\rangle_\textit{VIII}$ defined in (\ref{eq:8spinSignal}) for the 13-$\mu\text{s}$ pulse and (b) simulations of $\left\langle 2 J_z \right\rangle_\textit{VIII}$ for the 7-$\mu\text{s}$ pulse as a function of $n$, the number of pulse applied.}
\end{figure}
\end{itemize}

\begin{figure}
\includegraphics[scale=0.5]{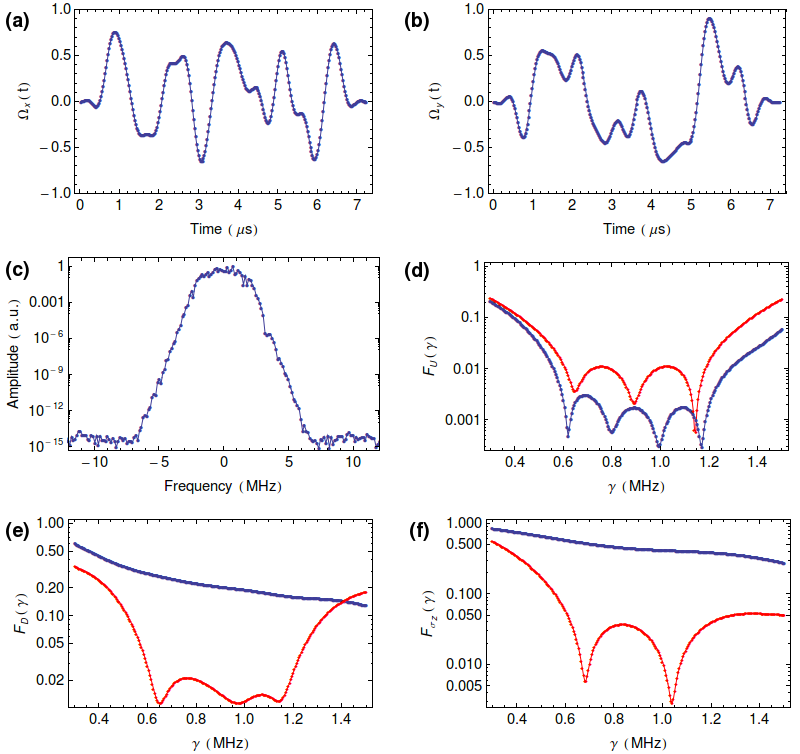}
\caption{\label{fig:7usPulse}(a) In phase amplitude $\Omega_x(t)$ of the 7-$\mu\text{s}$ pulse. (b) Quadrature amplitude $\Omega_y(t)$ of the 7-$\mu\text{s}$ pulse. (c) Absolute value of the Fourier transformed pulse centered at the carrier frequency. (d) Unitary metric $F_U(\gamma)$ defined in (\ref{eq:unitaryMetric}) for the 7-$\mu\text{s}$ pulse (blue) and for the 13-$\mu\text{s}$ pulse (red). (e) Dipolar metric $F_D(\gamma)$ defined in (\ref{eq:dipolarMetric}) for the 7-$\mu\text{s}$ pulse (blue) and for the 13-$\mu\text{s}$ pulse (red). (f) $\sigma_z$ metric $F_{\sigma_z}(\gamma)$ defined in (\ref{eq:chemicalMetric}) for the 7-$\mu\text{s}$ pulse (blue), and for the 13-$\mu\text{s}$ pulse (red).}
\end{figure}

The target for the pulse search without the AHT terms was $\Phi \left(\lbrace \Omega_x^{(i)}, \Omega_y^{(i)} \rbrace \right) = \frac{1}{14} \sum_{\gamma_i \in \Gamma} F_U \left(\lbrace \Omega_x^{(i)}, \Omega_y^{(i)} \rbrace, \gamma_i \right)$, $\Gamma$ being the same set of fourteen $\gamma$ values used in Section \ref{sec:PulseSearch}. As before, we forced the beginning and the end of the pulse to have zero amplitude and used the same transfer function as for the search in Section \ref{sec:PulseSearch}. This time we terminated the minimization only once the performance of the pulse without any dipolar effects nor resonance offsets mimicked the performance of the 13-$\mu\text{s}$ pulse. To determine that we evaluated $\left\langle \sigma_z \right\rangle_\textit{I}(n)$, given by (\ref{eq:1spinSignal}), for $250$ consecutive applications of either pulse, the results are displayed in Fig. \ref{fig:1SpinSim}. We use this kind of termination condition rather than matching the $F_U(\gamma)$ curves for the pulses to ensure that the eight spin simulations comparing the pulses will highlight errors resulting from evolution under the dipolar and $\sigma_z$ Hamiltonians rather than incoherence from the Rabi dispersion. Analogously to the search in Section \ref{sec:PulseSearch}, we increased the pulse length until we found a pulse satisfying our criteria. The pulse found was $7$-$\mu\text{s}$ long with $N=292$. Its waveform and metrics are presented in Fig. \ref{fig:7usPulse}. It can be seen that while the unitary metric $F_U(\gamma)$ for the shorter pulse is actually somewhat better than the one for the averaging pulse, its dipolar and $\sigma_z$ metrics $F_D(\gamma)$ and $F_{\sigma_z}(\gamma)$ are substantially worse.

Finally, we evaluated $\left\langle 2 J_z \right\rangle_\textit{VIII}$ given by (\ref{eq:8spinSignal}) for the pulses presented in Figs. \ref{fig:13usPulse} \& \ref{fig:7usPulse} for 250 back-to-back applications of the pulses (Fig. \ref{fig:8SpinSim}). It can be seen that the pulse without AHT terms performs about $7.1$ times worse despite being roughly half as long. Importantly, the performance of the 7-$\mu\text{s}$ pulse is sufficiently bad that it would not have enabled the experiments discussed in the main text. Separate simulations revealed that the signal decay for the 7-$\mu\text{s}$ pulse was mostly due to dipolar evolution; the removal of the resonance offsets increased the effective $e^{-1}$ time by a factor of $1.3$, while the removal of resonance offsets for the 13-$\mu\text{s}$ pulse increased the $e^{-1}$ time by a factor of $1.5$.